\documentclass[pre,aps,showpacs,twocolumn]{revtex4}
\usepackage{amsmath}
\usepackage{epsfig}          
\begin{document}
\title{Depinning of semiflexible polymers in ${\bf (1+1)}$ dimensions}
\author{P. Benetatos$^1$ and E. Frey$^{1,2}$}
\affiliation{$^1$Hahn-Meitner-Institut, Abteilung Theoretische Physik,
Glienicker Str. 100, D-14109 Berlin, Germany\\
Fachbereich Physik, Freie Universit\"at Berlin, Arnimallee 14, D-14195 Berlin, Germany}


\date{\today}

\begin{abstract}

We present a theoretical analysis of a simple model of the depinning of an anchored semiflexible polymer from a fixed planar substrate in $(1+1)$ dimensions. We consider a polymer with a discrete sequence of pinning sites along its contour. Using the scaling properties of the conformational distribution function in the stiff limit and applying the necklace model of phase transitions in quasi-one-dimensional systems, we obtain a melting criterion in terms of the persistence length, the spacing between pinning sites, a microscopic effective length which characterizes a bond, and the bond energy. The limitations of this and other similar approaches are also discussed. In the case of force-induced unbinding, it is shown that the bending rigidity favors the unbinding through a ``lever-arm effect''.

\end{abstract}

\maketitle

\section{Introduction}

For a broad category of physical problems, a free polymer is characterized by two lengths: the total contour length ($L$) and the persistence length ($L_p$) which is the correlation length of the tangent unit vector along its contour and is proportional to its bending rigidity. When the persistence length is much smaller than the total length, the polymer is said to be flexible and can be treated as a random walk. When the two lengths are of the same order, the polymer is said to be semiflexible. Some of the most important biopolymers belong to the latter class. For example, the structural elements of the cytoskeleton are microtubules, actin filaments, and intermediate filaments with persistence lengths of the order of $6mm$ \cite{MHow}, $17\mu m$ \cite{How}, and $2\mu m$ \cite{Bray} respectively. Although DNA filaments usually have a total length greater than the persistence length ($L_p\approx 50nm$), the latter is long enough to affect their elastic properties.\cite{MS} Obvious biological relevance and inherent theoretical challenges have sparked great interest in the statistical mechanics of semiflexible polymers in recent years.\cite{review}

A theoretical analysis of the unbinding of semiflexible polymers from fixed surfaces or interfaces (adsorption-desorption transition) or of two semiflexible strands from each other is a particularly tricky problem. The main reason is that sharp phase transitions in statistical mechanics occur only in the thermodynamic limit and the thermodynamic limit of semiflexible polymers is ambiguous. If we keep the persistence length fixed and take the total contour length to infinity, we obtain a flexible polymer. If we take the persistence length to infinity keeping the total length fixed, we obtain a rigid rod without any fluctuations. There have been several studies of this subject over the past few years.\cite{MHL,GB,GS,KS1,KS2,Step,BL,MKS} In all of those works the polymer binds to a potential well which {\it continuously} extends over the surface (or interface). In this article, we consider a simple model where a weakly bending semiflexible polymer in $(1+1)$ dimensions is bound to a fluid surface (or interface) through a {\it discrete} sequence of regularly placed pinning sites (``sticky points'') along its length. Such a model resembles the physical situation where an actin filament binds to a membrane through anchoring proteins.\cite{CPC} The discreteness of the binding sites allows us to employ a different kind of thermodynamic limit which avoids the inconsistencies that appear in some previous works. For fixed total length and persistence length (with $L\ll L_p$), we take the density of binding sites to infinity under the constraint that the probability of finding such a site inside the binding region remains constant. Thus we obtain a second order unbinding transition and a melting temperature which is a function of of the persistence length, the spacing between pinning sites, a microscopic length which characterizes a bond, and the bond energy. The article is organized as follows: In Sec. II  we calculate the probability of finding a binding site of the polymer inside a small (microscopic) region which characterizes a bond. We then consider the simplest version of our model which is a filament with only one pinning site. In Sec. III we use a necklace model \cite{HF,Fisher} type of approach which yields the thermal depinning  transition. In Sec. IV we discuss the effect of the bending rigidity on the force-induced unbinding of semiflexible polymers in the stiff limit. Finally, in Sec. V we demonstrate the subtleties of this problem comparing our model with other approaches and we present our conclusions.

\section{Conformational Probability - Formalism}

A widely used model which captures much of the physics of semiflexible polymers (except for their self-avoidance) is the worm-like chain (WLC)\cite{STY}  where the polymer is considered to be a continuous inextensible curve ${\bf r}(s)$ parametrized by the arc length $s$ measured along its contour from a fixed end. The effective free energy of a particular conformation depends only on the bending (curvature) and is given by
\begin{equation}
\label{WLCHam}
{\cal H} = \frac{\kappa}{2}\int_0^L ds \Big [ \frac{\partial {\bf t}(s)}{\partial s} \Big ]^2 \;,
\end{equation}
where ${\bf t}(s)={\partial {\bf r}(s)}/{\partial s}$ is the tangent unit vector of the curve ${\bf r}(s)$ and $\kappa$ is the bending rigidity which is related to the persistence length via $L_p=2\kappa/k_BT$ (in two dimensions). 

The orientational probability distribution function for free semiflexible chains having an initial tangent vector ${\bf t}(0)={\bf t}_0$ and a final tangent vector  ${\bf t}(L)={\bf t}_L$ is given by the path integral
\begin{eqnarray}
\label{Green}
& & G({\bf t}_L, L | {\bf t}_0, 0)= \nonumber\\
& & {\cal N} \int_{{\bf t}(0)={\bf t}_0}^{{\bf t}(L)={\bf t}_L}{\cal D}[{\bf t}(s)]\delta[|{\bf t}(s)|-1]\exp{\Big[-\frac{\cal H}{k_BT}\Big]}\;,
\end{eqnarray}
where the integration is over all fluctuating ``paths'' ${\bf t}(s)$ subject to the fixed boundary conditions and the inextensibility constraint $|{\bf t}(s)|=1$. ${\cal N}$ is a normalization constant. There is a formal analogy between the classical statistical mechanics of a semiflexible polymer and the quantum statistical mechanics of a rigid rotator.\cite{STY} If we make the correspondence ${\kappa}\rightleftharpoons I $, $k_BT\rightleftharpoons\hbar$, and $L\rightleftharpoons \beta\hbar$ in Eqs. (\ref{WLCHam},\ref{Green}), we notice that $G({\bf t}_L, L | {\bf t}_0, 0)$ corresponds to the density matrix element, in the angle representation, of a quantum rigid rotator with moment of inertia $ I $ and inverse temperature $\beta$. As in the case of a density matrix,\cite{Feynman} the angular probability distribution function of a free semiflexible polymer satisfies a Schr\"odinger equation in imaginary time: 
\begin{equation}
\label{Schr}
\frac{\partial G}{\partial s}=\frac{1}{L_p}\frac{{\partial}^2 G}{\partial \theta^2}\;,
\end{equation}
where $\theta(s)$ is the angle between ${\bf t}(s)$ and a fixed reference axis.\cite{ft1} In order to obtain the complete distribution function which in addition to the tangent vector also includes the position vector, $G({\bf r}_s, {\bf t}_s, s|{\bf r}_0, {\bf t}_0, 0)$, we have to replace the $s$-derivative in the lhs of Eq. (\ref{Schr}) by the ``convective'' derivative ${\partial}_s + {\bf t}\cdot\nabla_{\bf r}\;$ along the polymer ``path'' ${\bf r}(s)$ with instantaneous position vector ${\bf r}$ and tangent vector ${\bf t}$.\cite{Gobush} In Cartesian coordinates, the equation reads
\begin{eqnarray}
\label{Schr2}
& &\Big[\frac{\partial}{\partial s} + \cos\theta \frac{\partial}{\partial x} + \sin\theta \frac{\partial}{\partial y} - \nonumber\\
& & \frac{1}{L_p}\frac{\partial^2}{\partial \theta^2}\Big]G(x_s, y_s, \theta_s, s|x_0, y_0, \theta_0, 0)=0
\end{eqnarray}
where $\theta$ is the local slope of the polymer with respect to the $x$-axis. 

In the weakly bending limit ($L\ll L_p$), $\;\theta\ll 1$ and we simplify Eq. (\ref{Schr2}) setting $\sin\theta \approx \theta$ and $ \cos\theta \approx 1$. Since we  are not interested in the longitudinal fluctuations of the polymer (along the $x$-axis), we integrate the complete probability distribution function over $x$ to obtain a simpler equation for the reduced probability distribution: 
\begin{eqnarray}
\label{Schr3}
\Big[\frac{\partial}{\partial s} + \theta \frac{\partial}{\partial y} - 
 \frac{1}{L_p}\frac{\partial^2}{\partial \theta^2}\Big]G(y_s, \theta_s, s|y_0, \theta_0, 0)=0 \;.
\end{eqnarray}
Using Fourier transformations \cite{Risken}, we solve Eq. (\ref{Schr3}) with the ``initial'' condition $\lim_{s\rightarrow 0}G(y_s, \theta_s, s|y_0, \theta_0, 0)= \delta(\theta-\theta_0)\delta(y-y_0)$ to get
\begin{eqnarray}
\label{prob}
& &G(y_s, \theta_s, s|y_0, \theta_0, 0)= \nonumber\\
& &\frac{\sqrt3}{2\pi}\frac{L_p}{s^2}\exp\Big\{-\frac{3L_p}{s^3}[(y-y_0-\theta_0 s)^2- \nonumber\\
& &s(y-y_0-\theta_0 s)(\theta-\theta_0)+\frac{1}{3}s^2(\theta-\theta_0)^2]\Big\}\;.
\end{eqnarray}
Apart from explicitly containing the persistence length $L_p\;$, Eq. (\ref{prob}) is identical with that obtained in Ref. \cite{GB}. The interpretation, however, is very different. In Refs. \cite{GB,MHL},  $\;G(y_s, \theta_s, s|y_0, \theta_0, 0)$ is interpreted as a {\it dimensionless} partition function {\it independent} of the persistence length $L_p$ which has been eliminated by rescaling $y$ and $\theta$. In those references, Eq. (\ref{prob}) is expected to be valid for large $s$ and it appears that $s$ is measured in units of an extra, ``monomer'' length. In contrast, we interpret it as a two-point conformational {\it probability distribution} valid only in the weakly bending limit ($s\ll L_p$). Notice that $G(y_s, \theta_s, s|y_0, \theta_0, 0)$ fulfils the three fundamental properties of a two-point probability distribution; its integral over $y_s$ and $\theta_s$ is $1$, it becomes a delta function when $s\rightarrow 0$ and it obeys the Chapman-Kolmogorov equation.\cite{ft2} The corresponding partition function differs from $G(y_s, \theta_s, s|y_0, \theta_0, 0)$ by a normalization fuctor (related to the measure of the path integral) which should have units of length in order to render it dimensionless. (It is similar to the phase volume element $2\pi \hbar$ used in the statistical mechanics of gases.) In the calculation of several quantities, this normalization factor is unimportant as it drops out. For this type of problems, $G(y_s, \theta_s, s|y_0, \theta_0, 0)$ itself can be considered as the partition function. However, as it will become clear below, the necklace model involves a sum over powers of the partition function and using a dimensionful quantity in its place would clearly be erroneous.

For fixed $y_0=\theta_0=0$, the mean square slope and transverse displacement  of the free end of a filament of length $L$ are $\langle\theta_L^2\rangle=2L/L_p$ and $\langle y_L^2\rangle=(2/3)L^3/L_p$ as can be easily calculated from Eq. (\ref{prob}). The {\it probability} of finding the free end within a very small range of slopes and transverse displacements ($-\delta<\theta_L<\delta$ and $-\epsilon<y_L<\epsilon$ with $0<\delta, \epsilon\ll 1$) is
\begin{eqnarray}
\label{prob2}
& &P(\delta, \epsilon, L, L_p)=\int_{-\delta}^{\delta}d\theta_L\int_{-\epsilon}^{\epsilon}dy_LG(y_L, \theta_L, L|0, 0, 0) \nonumber\\
& &\approx \frac{\sqrt 3}{2\pi}\frac{L_p}{L^2}B\;,
\end{eqnarray}
where $B\equiv 4\delta\epsilon$ and the approximation holds for $B\ll ({\sqrt3}/2\pi)L^2/L_p\;$. The {\it partition function} $Z(\delta, \epsilon, L, L_p)$ of a polymer which is constrained so that $y_0=\theta_0=0$ and $-\delta<\theta_L<\delta\;$, $-\epsilon<y_L<\epsilon$ while it is unconstrained in the longitudinal direction is related to the probability $P(\delta, \epsilon, L, L_p)$ via $Z(\delta, \epsilon, L, L_p)=
Z_f(L, L_p)P(\delta, \epsilon, L, L_p)$, where $Z_f(L, L_p)$ is the partition function of a free filament. The latter has the property $Z_f(L_1+L_2, L_p)=Z_f(L_1, L_p)Z_f(L_2, L_p)$ and will be neglected as it is not going to affect any of the observable quantities we are interested in.

The probability of finding both the free end and the point at the middle confined within a very small range of slopes and transverse displacements is 
\begin{eqnarray}
\label{prob3}
& &\int_{-\delta}^{\delta}d\theta_L\int_{-\epsilon}^{\epsilon}dy_LG(y_L, \theta_L, L|y_{L/2}, \theta_{L/2}, L/2)\times   \nonumber\\
& &\int_{-\delta}^{\delta}d\theta_{L/2}\int_{-\epsilon}^{\epsilon}dy_{L/2}G(y_{L/2}, \theta_{L/2}, L/2|0, 0, 0)\approx \nonumber\\
& &\Big(\frac{\sqrt 3}{2\pi}\frac{L_p}{L^2}B\Big)^2\;.
\end{eqnarray}
We shall use this factorization in the calculation of the partition function of our model.

We now consider the toy system of a weakly bending semiflexible polymer with its endpoints ($s=0$ and $s=L$) bound and a pinning site in the middle ($s=L/2$). A ``bound site'' in our model is defined as a point of the polymer which is constrained to fluctuate within a {\it microscopically} small range of slopes and transverse displacements which is characterized by the effective length $B$ as defined above but is free to fluctuate in the longitudinal direction. The latter situation is physically realized in the case of a fluid substrate where the ``sticky points'' are free to move along a $1d$ track (membrane). A ``pinning site'' is defined as a point on the polymer which is energetically favorable to be bound with an associated bond energy $J\;$ ($J>0$). The partition function of this system with one pinning site is
\begin{equation}
\label{Z_1}
Z_1=[{\cal G}(L/2)]^2v + {\cal G}(L) - [{\cal G}(L/2)]^2\;,
\end{equation}
where 
\begin{equation}
\label{stw}
{\cal G}(l)=\frac{\sqrt 3}{2\pi}\frac{L_p}{l^2}B
\end{equation}
is the conformational statistical weight of a polymer segment of contour length $l$ whose endpoints are bound and $v\equiv \exp(J/k_BT)\;$. The third term in Eq. (\ref{Z_1}) is the ``counterterm'' needed to prevent doublecounting of conformations; the conformations associated with $[{\cal G}(L/2)]^2$ have already been included in ${\cal G}(L)$.

The average fraction of intact bonds is
\begin{eqnarray}
\label{fraction}
Q=\frac{\partial \ln Z_1}{\partial \ln(v-1)}\;.
\end{eqnarray}
This is a general expression valid for any number of pinning sites provided that we replace $Z_1$ with the corresponding $Z_N$ and we divide the rhs by $N$. The calculation of $Z_N$ for $N\gg 1$ is the aim of Sec. III.

\section{Thermal Depinning Transition}

The partition function of a weakly bending semiflexible polymer with its endpoints ($s=0$ and $s=L$) bound and $N$ pinning sites regularly distributed along its length formally reads
\begin{equation}
\label{Zformal}
Z_N=\sum_{n=0}^{N}v^n{\cal P}(n)\;,
\end{equation}
where ${\cal P}(n)$ is the probability of a conformation with exactly $n$ bonds ({\it but not} $n+1$ or $n+2$ or... $N$ bonds). For example, in the case of $N=2$, ${\cal P}(2)=[{\cal G}(L/3)]^3$, $\;{\cal P}(1)=2{\cal G}(L/3)\{{\cal G}(2L/3)-[{\cal G}(L/3)]^2\}$, and $\;{\cal P}(0)={\cal G}(L)-2{\cal G}(L/3)\{{\cal G}(2L/3)-[{\cal G}(L/3)]^2\}-[{\cal G}(L/3)]^3\;$. Collecting terms, we obtain: $Z_2=[{\cal G}(L/3)]^3(v-1)^2+2[{\cal G}(L/3)]{\cal G}(2L/3)(v-1)+{\cal G}(L)\;$.


\begin{figure}
\begin{center}
\label{cartoon}
\leavevmode
\hbox{%
\epsfxsize=3.2in
\epsffile{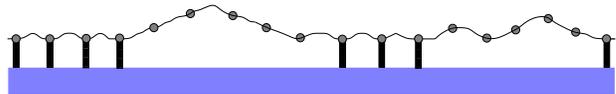}}
\end{center}
\caption{A cartoon picture of a polymer conformation with a ``chain'' of length $3L_m$, a ``bubble'' of length $6L_m$, a ``chain'' of length $2L_m$ and a ``bubble'' of length $6L_m$. The dots represent pinning sites and the vertical black lines represent anchoring proteins which bind the filament to the substrate. The weakly bending limit allows us to neglect any direct (hard wall) interaction with the substrate. }
\end{figure}


Let us define a ``bubble'' as a polymer segment with only its ends bound. The minimum length of a bubble is $L_m=L/(N+1)\;$ and the maximum length is equal to the total contour length of the filament $L$. Notice that in our model a bubble will always have length $\leq L\ll L_p$. A ``chain'' is defined as a sequence of minimal bubbles (of length $L_m\;$). Fig. 1 provides a pictorial definition of ``bubbles'' and chains. The partition function $Z_N$ is a sum which consists of all products of the form $f(m_1)g(n_1)f(m_2)g(n_2)...f(m_k)$ where 
\begin{equation}
\label{f}
f(m)=\frac{\sqrt 3}{2\pi}\frac{L_p B}{L_m^2 m^2}
\end{equation}
is the statistical weight of a bubble of length $mL_m$,
\begin{equation}
\label{g}
g(n)=[f(1)]^n(v-1)^{n+1}
\end{equation}
is the statistical weight of a chain of length $nL_m$,
and
\begin{equation}
\label{constraint}
m_1+n_1+m_2+n_2+...+m_k=N+1, \quad 0<k<\Big[\frac{N+1}{2}\Big],
\end{equation}
where $[(N+1)/2]$ is the integer part of $(N+1)/2\;$. We have:
\begin{eqnarray}
\label{Z_Nsum}
Z_N&=&\sum_{m=1}^Nf(m)f(N+1-m)(v-1)+\nonumber\\
& & \sum_{k=1}^{[(N+1)/2]}\sum_{\{n_j\}}\sum_{\{m_j\}}\frac{\prod_{j=1}^k f(m_j)g(n_j)}{g(n_k)}\;.
\end{eqnarray}
The first sum in Eq. (\ref{Z_Nsum}) represents configurations with only one bond and we shall denote it by $D_N\;$. The curly braces indicate that the sums must satisfy the constraint of Eq. (\ref{constraint}). As $N$ increases, calculating the combinatorial factors becomes an impossible task. That is why we use a standard trick and incorporate the constraint in the partition function via a Kronecker delta \cite{WM}:
\begin{eqnarray}
\label{Z_Nsum2}
& &Z_N=D_N+
\sum_{k=1}^{[(N+1)/2]}\sum_{n=1}^{\infty}\sum_{m=1}^{\infty}\delta[N-\sum_{j=1}^{k}(n_j+m_j)]\times\nonumber\\
& &\exp\{\beta[N-\sum_{j=1}^{k}(n_j+m_j)]\}\frac{\prod_{j=1}^k f(m_j)g(n_j)}{g(n_k)}\;,
\end{eqnarray}
where $\sum_{n=1}^{\infty}\sum_{m=1}^{\infty}\equiv\sum_{n_1=1}^{\infty}...\sum_{n_k=1}^{\infty}\sum_{m_1=1}^{\infty}...\sum_{m_k=1}^{\infty}$ and the auxiliary real parameter $\beta$ has been introduced to make sure that the partition function converges in later steps of the calculation.

We now introduce a complex representation of the Kronecker delta which yields:
\begin{eqnarray}
\label{Z_Nsum3}
& &Z_N=D_N+\frac{1}{2\pi}\int_0^{2\pi}d\theta\exp[N(\beta+i\theta)]\times\nonumber\\
& &\sum_{k=1}^{[(N+1)/2]}\frac{\prod_{j=1}^k \Phi_j \Psi_j}{\Phi_k}\;,
\end{eqnarray}
where 
\begin{equation}
\label{Phi}
\Phi_j=\sum_{n_j=1}^{\infty}g(n_j)z^{n_j}\;,
\end{equation}
\begin{equation}
\label{Psi}
\Psi_j=\sum_{m_j=1}^{\infty}f(m_j)z^{m_j}\;,
\end{equation}
with $z\equiv\exp[-(\beta+i\theta)]\;$.

Since both $\Phi_j$ and $\Psi_j$ are independent of $j$, $\prod_{j=1}^k\Phi_j\Psi_j=(\Phi \Psi)^k$, and since we are interested in a very large number of pinning sites (thermodynamic limit, $N\to\infty$), we approximate $\sum_{k=1}^{[(N+1)/2]}(\Phi \Psi)^k\approx\Phi \Psi/(1-\Phi \Psi)\;$. Using analytic continuation, we transform the integral over $\theta$ to a contour integral over the complex ``fugacity'' $z$ where the contour encircles the origin $z=0$ in the counterclockwise direction once and we obtain:
\begin{equation}
\label{Zfinal}
Z_N=D_N+\frac{1}{2\pi i}\oint_C dz z^{-N-1} \frac{\Psi}{1-\Phi\Psi}\;,
\end{equation}
where
\begin{equation}
\label{Phi_2}
\Phi=\frac{f(1)(v-1)^2z}{1-f(1)(v-1)z}
\end{equation}
and
\begin{equation}
\label{Psi2}
\Psi=\frac{\sqrt 3}{2\pi}\frac{L_p}{L_m^2}BL_2(z)
\end{equation}
with $L_2(z)$ being Euler's dilogarithm function.\cite{Erd}
The integrand in Eq. (\ref{Zfinal}) has three singularities: a pole of order $N+1$ at $z=0$; a simple pole at the solution $z_0$ of equation 
\begin{equation}
\label{pole}
\Psi(z_0)\Phi(z_0)=1 ; 
\end{equation}
and a branch cut along the positive real axis starting at $z=1$ due to $L_2(z)\;$. The contour $C$ encircles only the singularity at the origin because of the assumptions that we had made in deriving Eq. (\ref{Zfinal}). That is, $|z|$ was chosen so that the series $\Phi$ and $\Psi$ converge and also $|\Phi\Psi|<1\;$. As shown in Fig. 2, the contour $C$ can be deformed into a contour which encircles only $z_0$ clockwise and a loop which goes around the branch cut and closes at infinity counterclockwise.\cite{Wiegel} At sufficiently low temperatures, $0<z_0<1\;$. In the thermodynamic limit, $D_N$ vanishes and the partition function is determined by the pole at $z_0$: $N^{-1}\ln Z_N \approx -\ln z_0\;$. 


\begin{figure}
\begin{center}
\label{complex plane}
\leavevmode
\hbox{%
\epsfxsize=3.2in
\epsffile{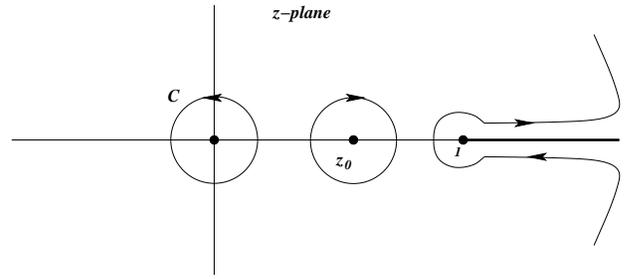}}
\end{center}
\caption{The complex $z$ plane with poles at $z=0$, $z=z_0$ and a branch cut along the positive real axis starting at $z=1$. The contour $C$ can be deformed to a contour around $z_0$ and a contour around the branch cut which closes at infinity.}
\end{figure}


As the temperature increases, $v$ decreases and it can be seen from Eq. (\ref{pole}) that $z_0$ is shifted to the right. In the thermodynamic limit, Eq. (\ref{fraction}) yields $Q=-[(v-1)/z_0](\partial z_0/\partial v)$. This implies that the average fraction of intact bonds, which is a physically observable quantity, monotonically decreases.\cite{ft2} The unbinding transition occurs when $z_0\to1\;$. Therefore, the unbinding criterion is $\Phi(1)\Psi(1)=1\;$. Given that $L_2(1)=\zeta(2)\approx 1.64$, where $\zeta(z)$ is Riemann's zeta function, we obtain
\begin{equation}
\label{transition}
\frac{L_pB}{L_m^2}[\exp(J/k_BT_c)-1]\approx 2\;.
\end{equation}
This is the main result of the article. The transition will be of second order because the derivative of $L_2(z)$ diverges logarithmically at $z=1$ and therefore the average fraction of intact bonds vanishes continuously at the critical temperature $T_c$. Notice that in order to obtain this transition we need the thermodynamic limit where $N\to \infty$ {\it and} $\frac{\sqrt 3}{2\pi}\frac{L_p}{L^2}N^2B$  which is the probability of finding a pinning site within the binding region remains constant $\ll 1\;$. This implies that $B\to 0$ which is consistent with a {\it microscopic} effective length that characterizes the bond. Of course, real systems will have a finite number of pinning sites and the transition will not be sharp. For a sufficiently high density of pinning sites, however, we would expect a clear crossover from a low temperature phase with most of the pinning sites bound to a high temperature phase with most of the pinning sites unbound which will be described by Eq. (\ref{transition}).

\section{Bending rigidity and force-induced unbinding}

In this Section we consider the force-induced unbinding of a weakly bending semiflexible polymer and we show that the bending rigidity facilitates the unbinding. If we apply a transverse force to the free end of a clamped semiflexible polymer, the effective free energy of Eq. (\ref{WLCHam}) changes by an extra term:
\begin{equation}
\label{fHam}
{\cal H}_f = \frac{\kappa}{2}\int_0^L ds \Big [ \frac{\partial {\bf t}(s)}{\partial s} \Big ]^2 - f\int_0^L ds \sin\theta(s) \;,
\end{equation}
where, as in Sec. II, $\theta(s)$ is the slope of the tangent vector with respect to the longitudinal direction. The partition function of this system is a path integral over all possible conformations. Slicing the $L$-length into $N$ segments each of length $a$ and using the small-angle approximation, we obtain: 
\begin{equation}
\label{HamN}
{\cal H}_f^N=\frac{\kappa}{2a}\sum_{i=1}^{N}(\theta_i-\theta_{i-1})^2 - fa\sum_{i=1}^{N}\theta_i\;.
\end{equation}
In this approximation, the path integral is Gaussian and can be easily calculated\cite{Feynman} yielding
\begin{equation}
\label{fZ}
Z_f=\exp\Big[\frac{f^2L^3}{3L_p(k_BT)^2}\Big]\;.
\end{equation}
The corresponding free energy is
\begin{equation}
\label{fF}
F=-\frac{f^2L^3}{3L_p(k_BT)}\;.
\end{equation}
Notice that this free energy is just minus the elastic energy of a cantilever spring, $U_{el}=f^2/2\chi$, whith spring constant $\chi=3\kappa/L^3\;$. The latter has been obtained in a linear response calculation in Ref. \cite{Klaus}.

Although the original ``Hamiltonian'' (Eq. \ref{fHam}) is extensive, the free energy $F$ of Eq. (\ref{fF}) is not (grows as $L^3$) because Eq. (\ref{fF}) is an approximation valid only in the short length scales of the weakly bending limit ($L\ll L_p$). Given that the free energy of the bound state is always extensive, this non-extensivity leads to a ``lever-arm effect'' in the force-induced unbinding. That is, for a long enough total length, the unbound state will be favorable having a lower free energy. We can estimate an upper bound for this ``critical'' length. If ${\cal F}_b$ is the free energy density of the bound state, the ``lever-arm'' critical length $L_l$ should satisfy the condition:
\begin{equation}
\label{lever}
{\cal F}_b L_l =\frac{f^2L_l^3}{3L_p(k_BT)}\;.
\end{equation}
If $L>L_l$, the transverse force $f$ will always unbind the polymer (in equilibrium). Using the model of Sec. III, we have ${\cal F}_b \approx ( k_B T/L_m)[J/k_BT-\ln(\sqrt3 L_p B/L_m^2 2 \pi)]\;$. It turns out that 
\begin{equation}
\label{L_l}
L_l\approx \sqrt3 \frac{1}{f}\Big[J- k_B T\ln\Big(\frac{\sqrt3 L_p B}{L_m^2 2 \pi}\Big)\Big]^{1/2}(\frac{L_p}{L_m})^{1/2}\;.
\end{equation}
For pulling forces of the order of picoNewton, persistence length of the order of $\mu m$, and binding free energy per $L_m$ of the order of $k_BT$, it turns out that $L_l\approx10^{-3}(L_p/L_m)^{1/2}L_p$ which is an indication of the relevance of the ``lever-arm effect'' to biopolymers.

This is a phenomenon related to the ``molecular leverage'' discussed in Ref. \cite{Bruinsma}. In both cases the bending rigidity facilitates the force-induced unbinding. The two phenomena, however, are different. We describe a situation where a bound state with a sequence of pinned sites becomes thermodynamically unstable when the system is long enough for the ``lever-arm'' to dominate the free energy  whereas Ref. \cite{Bruinsma} presents an estimate of the torque induced force exerted on a {\it single} ligand-receptor pair which turns out to be much stronger compared to that applied in traction.

\section{Discussion and Conclusions}

As we mentioned in the Introduction, the unbinding of semiflexible polymers is a particularly tricky problem because of the ambiguity of the relevant thermodynamic limit and also because of a lack of exact solutions of the WLC model with a binding potential. Refs. \cite{Step,KS1,KS2} actually deal with the unbinding of {\it flexible} polymers (with $L\gg L_p$) and consider the effect of the bending rigidity on the conformational properties of the adsorbed (low-temperature) phase. In these works, the bending rigidity enters as a perturbation to the flexible (Gaussian) chain and the inextensibility is absent. The early references \cite{MHL,GB} use a somewhat inconsistent approach where the scaling behavior of a weakly bending stiff filament (valid only for $L \ll L_p$) is artificially extended to apply to any length. Another serious drawback of this approach is that it yields results which appear to be independent of the persistence length while, in principle, they should not. The idea is to solve Eq. (\ref{Schr3}) for large $L$ with a binding potential using scaling Ans\"atze and then invoke the necklace model to predict the order of the unbinding transition from the scaling behavior of $G$. The details of the necklace model (``bubbles'', ``chains'', partition function, etc.), however, are not worked out. Ref. \cite{GS} employs a discrete model for stiff filaments where an extra (``monomer'') length is introduced and turns out to be relevant for  the unbinding transition. It also proposes an energy/entropy melting criterion which applies to adsorbed phases similar to those discussed in \cite{Step,KS1}. Ref. \cite{BL}, using a RG treatment, demonstrates the relevance of an orientation-dependent interaction field for the unbinding transition. The RG flow, however, implies a thermodynamic limit which carries on the inconsistencies of Refs. \cite{MHL,GB}. Ref. \cite{MKS} models a semiflexible polymer as a directed self-avoiding random walk and it reiterates the inconsistencies of Refs. \cite{MHL,GB} because the unbinding transition occurs at the thermodynamic limit of an infinitely long walk (infinitely longer than the persistence length) where one would normally expect to recover the behavior of a flexible chain.

In conclusion, applying the necklace model we have obtained a criterion for the depinning of anchored semiflexible polymers in the weakly bending (stiff) limit. This model has been extensively used to study the unbinding of flexible polymers \cite{PS,WM,Wiegel,KPM}. This is its first detailed application to the unbinding of semiflexible polymers. A general and rigorous theoretical treatment of the unbinding of semiflexible polymers of arbitrary total length and persistence length in the presence of an arbitrary binding potential has not yet been achieved. Our model suggests an alternative way to consider the thermodynamic limit for this system and straightens out several misconceptions of previous studies. 

We have also shown how the bending rigidity facilitates the force-induced unbinding of semiflexible polymers in the weakly bending limit. We have estimated a critical length as a function of the pulling force, the binding free energy density, the persistence length and the density of pinning sites above which the polymer acts as a lever-arm and unbinds.

\vspace{0.2in}
We thank T. Franosch, G. Gompper, M. C. Marchetti and A. Parmeggiani for useful comments and discussions.


\begin{thebibliography}{30}
\bibitem{ft2}
Not always can a path integral like that of Eq. (\ref{Green}) be interpreted as a two-point probability distribution function. For example, it can be shown that the solution of Eq. (\ref{Schr}) modified by an external stretching tension cannot fulfil the normalization requirement and the other two properties at the same time.
\bibitem{ft2}
The temperature dependence of the persistence length ($L_p$) does not alter this monotonicity.
\end{thebibliography}
\end{document}